# Searching the weakest link: Demagnetizing fields and magnetization reversal in permanent magnets

J. Fischbacher[1], A. Kovacs[1], L. Exl[2,3], J. Kühnel[4], E. Mehofer[4], H. Sepehri-Amin[5], T. Ohkubo[5], K. Hono[5], T. Schrefl[1]

[1] Center for Integrated Sensor Systems, Danube University Krems, 2700 Wiener Neustadt, Austria

[2] Faculty of Physics, University of Vienna, 1090 Vienna, Austria

[3] Faculty of Mathematics, University of Vienna, 1090 Vienna, Austria

[4] Faculty of Computer Science, University of Vienna, 1090 Vienna, Austria

[5] Elements Strategy Initiative Center for Magnetic Materials, National Institute for Materials Science, Tsukuba 305-0047, Japan

**Abstract**

Magnetization reversal in permanent magnets occurs by the nucleation and expansion of reversed domains. Micromagnetic theory offers the possibility to localize the spots within the complex structure of the magnet where magnetization reversal starts. We compute maps of the local nucleation field in a $Nd_2Fe_{14}B$ permanent magnet using a model order reduction approach. Considering thermal fluctuations in numerical micromagnetics we can also quantify the reduction of the coercive field due to thermal activation. However, the major reduction of the coercive field is caused by the soft magnetic grain boundary phases and misorientation if there is no surface damage.

**1. Introduction**

With the rise of sustainable energy production and eco-friendly transport there is an increasing demand for permanent magnets. The generator of a direct drive wind mill requires high performance magnets of 400 kg/MW power; and on average a hybrid and electric vehicle needs 1.25 kg of high end permanent magnets [1]. Modern high-performance magnets are based on $Nd_2Fe_{14}B$. These magnets have a high energy density product which means that the magnets can be small and still create a sufficiently large magnetic field. One weak point of $Nd_2Fe_{14}B$ is the relatively low Curie temperature as compared to SmCo based magnets. As a consequence, the coercive field of $Nd_2Fe_{14}B$ drops rapidly with increasing temperature. To enhance the operating temperature of $Nd_2Fe_{14}B$ magnets heavy rare earth elements are added. The anisotropy field of $(Nd,Dy)_2Fe_{14}B$ is higher than that of $Nd_2Fe_{14}B$ and therefore the Dy containing magnet can be operated at higher temperature. Production techniques that increase the Dy concentration near the grain boundary [2,3] reduce the share of heavy rare earths. In these magnets the anisotropy field is locally enhanced near the grain boundary which suppresses the formation of reversed domains [4,5]. Similarly, an enhancement of the coercive field has been achieved by Nd-Cu grain boundary diffusion [6].

The enhancement of coercivity by modification of the region next to the grain boundaries is a clear indication that magnetization reversal in permanent magnets is induced by the nucleation of a reversed domain near the grain surface. In Nd₂Fe₁₄B based permanent magnets weakly ferromagnetic grain boundary phases [7] act as nucleation sites [8] if there exist no grains with reversed magnetic domains at remanence. In addition to the presence of the soft magnetic grain boundary phase, magnetization reversal is facilitated by the local demagnetizing field. These fields obtain their highest values near the edges and corners of the polyhedral grain structure [9]. Traditionally the effect of defects where the intrinsic properties are different from the bulk and the effect of the demagnetizing field on coercivity is expressed as [10]

$$H_c(T) = \alpha H_N(T) - N_{eff} M_s(T). \tag{1}$$

Here $H_N = 2K_1/(\mu_0 M_s)$ is the ideal nucleation field. The constant $\mu_0$ is the permeability of vacuum. The anisotropy constant and the magnetization are denoted by $K_1$ and $M_s$. $H_N$ and $M_s$ are intrinsic magnetic properties and depend on the composition of the magnetic phase and on the temperature $T$. On the other hand, the coercive field, $H_c$, strongly changes with the microstructure of the magnet. Therefore, many researchers refer to $\alpha$ and $N_{eff}$ as microstructural parameters. The parameter $\alpha$ gives the reduction of the coercive field by soft magnetic defects and misorientation of the anisotropy axes with respect to the applied field direction; the parameter $N_{eff}$ describes the reduction of the coercive field owing to the self-demagnetizing field. It can be regarded as a local, effective demagnetization factor.

Equation (1) describes the influence of the microstructure on the coercive field. Temperature effects are included through the temperature dependence of $K_1(T)$, $M_s(T)$. A second mechanism how temperature influences the coercive field are thermal fluctuations on the macroscopic scale. These fluctuations may drive the system over a finite energy barrier. Hysteresis in a non-linear system like a permanent magnet results from the path formed by subsequently following local minima in an energy landscape constantly changed by a varying external field [11]. With increasing opposing field, the energy barrier that separates the system from the reversed state decreases. The critical field at which the energy barrier vanishes is the switching field of the magnet [12]. Switching at finite temperature can occur at non-zero energy barrier. If the system can escape over the energy barrier within the measurement time switching will occur. In permanent magnets it is assumed that the system can escape an energy barrier of 25 $k_B T$ within one second [13]. Therefore, to include the reduction of coercivity by thermal activation equation (1) can be rewritten as [14,15]

$$H_c(T) = \alpha H_N(T) - N_{eff} M_s(T) - H_f(T). \tag{2}$$

The thermal fluctuation field, $H_f$, can be expressed in terms of the activation volume $v$ [13]

$$H_f(T) = \frac{25\, k_B T}{\mu_0 M_s v} \tag{3}$$

Here $k_B = 1.38 \times 10^{-23}$ J/K is the Boltzmann constant. In this work we will quantify the influence of demagnetizing fields and thermal activation on the reduction of the coercive field with respect to the ideal nucleation field using micromagnetic simulations. The paper is organized as follows.

1. We will use equation (1) and analyze micromagnetic results for $H_c(T)$ to show how $\alpha$ and $N_{eff}$ changes with the microstructure.
2. We will apply a simplified micromagnetic model based on the local demagnetizing field near the grain boundaries to get a deeper insight on how demagnetizing effects reduce coercivity.
3. We will calculate the thermally activated switching to locate the weakest spot where magnetization reversal is initiated within a complex grain structure.

For the computations of $\alpha$ and $N_{eff}$ (see 1 above) we used a finite element micromagnetic solver [16]. The hysteresis loop is computed by minimization of the Gibbs free energy for decreasing external field. At all surfaces the mesh size is 2.4 nm. The simplified micromagnetic model (see 2 above) is based on a method for computing the demagnetizing field from surface charges [17] and a method for computing the switching field as function of field angle in the presence of defects [18]. Thermally activated switching (see 3 above) is computed using a modified string method [19,20]. The finite element meshes are created with Tetgen [21]. At all surfaces the mesh size is 2.4 nm.

## 2. Results

*2.1 Microstructural parameters and the demagnetizing field*

We computed the coercive field for a magnet consisting of equi-axed, platelet shaped, or columnar grains. The microstructure used for the simulations is shown in Figure 1. The grains are perfectly aligned. The edge length of the cube forming the magnet is 200 nm. The volume fraction of the grain boundary is 26 percent. Its thickness is 3.8 nm. For the grains we used the intrinsic magnetic properties of $Nd_2Fe_{14}B$ as function of temperature [22]. We performed two sets of simulations. In set 1 the grain boundary phase was non-magnetic; in set 2 the grain boundary phase was assumed to be weakly ferromagnetic. We set the magnetization of the grain boundary phase to 1/3 of the value for $Nd_2Fe_{14}B$, that is $M_{s,GB} = M_s/3$. The same material as in the grain boundary phase is used as surface layer with a thickness of 1.9 nm that covers the magnet. The purpose of this layer is to mimic surface damage. Please note that we change the exchange constant of all phases according to $A(T) = cM_s^2(T)$, where the factor $c$ is determined from the $Nd_2Fe_{14}B$ values at $T$ = 300 K ($\mu_0 M_s$ = 1.6 T and $A$ = 8 pJ/m). We apply the finite element method to compute the magnetization as function of the external field, $M(H_{ext})$. At each step of the external field the micromagnetic energy is minimized using a modified non-linear conjugate gradient method [16]. To compute the microstructural parameters, we applied the same procedure as usually done in experiments. We plotted $H_c(T)/M_s(T)$ as function of $H_N(T)/M_s(T)$ and fitted a straight line. Table 1 gives the microstructural parameters for the different structures. At all surface the mesh size was forced to be 2.4 nm.

Table 1. Microstructural parameter for magnets made of equi-axed, platelet shaped, and columnar grains.

| Shape of grains | Grain boundary phase | $\alpha$ | $N_{eff}$ |
|---|---|---|---|
| columnar | non-magnetic | 0.88 | 0.79 |
| equi-axed | non-magnetic | 0.88 | 0.87 |
| platelet | non-magnetic | 0.88 | 0.91 |
| columnar | weakly ferromagnetic | 0.45 | 0.10 |
| equi-axed | weakly ferromagnetic | 0.47 | 0.27 |
| platelet | weakly ferromagnetic | 0.51 | 0.43 |

The results clearly reflect the shape anisotropy of the grains. For both sets of simulations the demagnetizing factors increase as we go from columnar, equi-axed to platelet shaped grains. When a ferromagnetic grain boundary phase is introduced, the parameter $\alpha$ is reduced approximately by a factor of ½. The corresponding reduction in $N_{eff}$ may be understood with reduced surface charge at the grain surfaces. For a uniformly magnetized grain the demagnetizing field arises from magnetic surface charges which are proportional to $(M_s-M_{s,GB})$. So far, the numerical results for $\alpha$ and $N_{eff}$ are consistent with the conventional interpretation of the role played by soft magnetic defects and shape anisotropy.

The value of α is smaller than 1 for perfect grains without any defect and non-magnetic grain boundary. This can be explained by the non-uniform demagnetizing field within the magnet. The finite angle of the total field, which is the sum of the external field, the demagnetizing field and the exchange field, with respect to the anisotropy axis causes magnetization reversal by nucleation and expansion of reversed domains.

*2.2 Embedded Stoner Wohlfarth model*

We now show that we can understand the switching of grains in a permanent magnet using the model order reduction approach which we call embedded Stoner Wohlfarth method. By comparing the simplified method with conventional micromagnetic computations, we show that the perpendicular component of the demagnetizing field plays a significant role for magnetization reversal.

First, we compute the demagnetizing field of the magnet. Assuming that each grain is uniformly magnetized we can evaluate the demagnetizing field as a sum of line integrals along the edges of the grains [17]. For field evaluation we developed a highly parallel algorithm to be used on graphics processors. The demagnetizing field, $\mathbf{H}_{demag}$, is evaluated at several points $\mathbf{x}_i$ which are located at a distance $d$ from the edges of the grains. Then we loop the external field. For each value of the external field we compute the total field

$$\mathbf{H}_{tot} = \mathbf{H}_{ext} + \mathbf{H}_{demag} + \mathbf{H}_x, \tag{4}$$

at all points $\mathbf{x}_i$. We denote the angle between $\mathbf{H}_{tot}$ and the anisotropy axis, with $\psi$. According to the Stoner-Wohlfarth theory [24], the switching field of a small particle with uniform magnetization depends on the angle between the particle's easy axis and the field [25]

$$H_{sw} = f(\psi)H_N, \ \ f(\psi) = \frac{1}{\left(\cos^{2/3}\psi + \sin^{2/3}\psi\right)^{3/2}}. \tag{5}$$

The minimum value of the external field for which $|\mathbf{H}_{tot}| > H_{sw}$ is the switching field of the magnet.

*Embedded Stoner Wohlfarth method:*
compute the demagnetizing field $\mathbf{H}_{demag}$ and set $h = |\mathbf{H}_{ext}| = 0$
loop over *h*:
    evaluate $\mathbf{H}_{tot}(h)$ at all points $\mathbf{x}_i$
    compute $\psi_i$ and $H_{sw,i}$
    if $|\mathbf{H}_{tot}| > H_{sw,i}$ for any *i*
        set $H_c = h$ and stop
    increase *h*

The total field given by (4) is the sum of the external field, $\mathbf{H}_{ext}$, the demagnetizing field, $\mathbf{H}_{demag}$, and the exchange field, $\mathbf{H}_x$. Near edges the demagnetizing field is tilted and no more antiparallel to the magnetization [9]. Indeed, the perpendicular component of the demagnetizing field will grow to infinity as one approaches the edge. This divergence of the magnetostatic field is compensated by the exchange field leading to a finite total field at any point near the edge [23]. The exchange field is defined as follows: $H_{x,\parallel} = A/d^2$ and $H_{x,\perp} = 0$. The symbols ∥ and ⊥ denote the component of the exchange field parallel and perpendicular to the anisotropy axes.

In other words, we place virtual Stoner-Wohlfarth particles into each grain. These virtual particles are located at a distance *d* from the edges. Then we compute the Stoner-Wohlfarth switching fields of each virtual particle. The lowest value of the external field that leads to switching of at least one virtual Stoner-Wohlfarth particle gives the coercive field. Numerical experiments showed that at $d = 1.2\sqrt{A/(\mu_0 M_s^2)}$ the results of the embedded Stoner Wohlfarth method coincide with the coercive field obtained from conventional micromagnetics [26].

Figure 2 compares the grain size dependence of the coercive field computed with the embedded Stoner Wohlfarth method and conventional micromagnetics. The switching field of a single, isolated $Nd_2Fe_{14}B$ grain was calculated as function of grain size. There is a reasonable good agreement between the two methods. The inset shows a map of the local switching field according to the embedded Stoner Wohlfarth (ESW) model for $h = 0$. The lowest values of $H_{sw}$ are found near the bottom edge of the grain. This is exactly the region where a reversed domain is nucleated in conventional micromagnetics.

*2.3 Thermal activation*

We choose a grain boundary diffused magnet for analysis of thermal activation in permanent magnets. In particular we will show that the spot at which magnetization reversal is initiated can be tuned by grain boundary engineering. Figure 3 shows the microstructure used for the simulations. The magnet consists of 64 grains. The average grain size is 57 nm. The grains are separated by a weakly ferromagnetic grain boundary phase with a thickness of 3 nm. We investigate thermally activated magnetization reversal for different levels of grain boundary diffusion as shown on the right-hand side of Figure 3. For the intrinsic magnetic properties we used the room temperature values for $Nd_2Fe_{14}B$ and $(Nd,Dy)_2Fe_{14}B$. The material properties used for the $Nd_2Fe_{14}B$ core were $K_1$ = 4.3 MJ/m³, $\mu_0 M_s$ = 1.61 T, $A$ = 7.7 pJ/m. For the shell material, $(Nd,Dy)_2Fe_{14}B$, we assumed $K_1$ = 5.17 MJ/m³, $\mu_0 M_s$ = 1.15 T, $A$ = 8.7 pJ/m. The magnetocrystalline anisotropy of the grain boundary phase was zero. Its magnetization and exchange constant were $\mu_0 M_s$ = 0.5 T, $A$ = 7.7 pJ/m. The average degree of misalignment of the anisotropy directions was 15 degrees.

We consider three samples: no Dy diffusion, only the surface grains are covered with a dysprosium containing shell, and all grains are covered with a Dy containing shell. For these three samples we quantify the effects that reduce the coercive field with respect to the ideal nucleation field of $Nd_2Fe_{14}B$ main phase. To do so we perform three simulations. We denote the coercive fields obtained by these simulations with $H_{c,1}$, $H_{c,2}$, and $H_{c,3}$, respectively.

(1) We compute the demagnetization curve with a conventional micromagnetic finite element solver [16] but we switch off the magnetostatic field. We obtain the magnetization reversal process without any demagnetizing effects. The difference between the ideal nucleation field and the coercive field, $\Delta_{\psi,\text{defect}} = H_N - H_{c,1}$, has to be attributed to misorientation and soft magnetic defects.

(2) We perform a classical micromagnetic simulation of the reversal process. Now magnetostatic interactions are fully taken into account. The difference $\Delta_{\text{demag}} = H_{c,1} - H_{c,2}$ is caused by demagnetizing effects.

(3) We compute the energy barrier for switching as function of the applied field, $E(H_{\text{ext}})$ using a modified string method. The critical field, $H_{c,3}$, at which $E(H_{\text{ext}}) = 25\,k_B T$ is the nucleation field at temperature $T$ for a measurement time of one second. The difference $\Delta_{thermal} = H_{c,2} - H_{c,3}$ is caused by thermal fluctuations.

In order to address thermal activation in micromagnetics, we follow the procedure outlined by Fischbacher et al. [27]. Thermal fluctuations drive the magnetization over an energy barrier of finite size. We computed the minimum energy path, the most likely path the system takes over an energy barrier, and the associated saddle point using a modified string method [28]. Our implementation of the string method uses energy minimization [19] and path truncation [26].

We quantify the effects that reduce the ideal nucleation field, $H_N$, in the magnets. The results are presented in Figure 4. In all three samples, the major reduction of the coercive field is caused by misorientation and the weakly soft magnetic grain boundary. The reduction by misorientation and the soft-magnetic grain boundary phase is about 50 percent of the ideal nucleation field for the Dy free magnet. It is reduced to about 40 percent of the ideal nucleation field for the fully Dy diffused sample. The reduction owing to demagnetizing effects is smaller than 10 percent. For our samples we find $0.07\, H_N > \Delta_{demag} \geq 0.03\, H_N$. The reduction of the coercive field by thermal fluctuations is slightly smaller.

## 3. Discussion

We showed that the switching field of polyhedral grains can be computed with a simplified model that takes into account the local demagnetization field and applies the Stoner Wohlfarth model locally at points close the edges of the grains. The good agreement of the embedded Stoner Wohlfarth model with conventional micromagnetics suggests that magnetization reversal is governed by strength and orientation of the demagnetizing field near the edges of the magnet. At the point where magnetization reversal starts, the angle between the total field and the anisotropy axis is greater than zero. This effective misorientation, which is caused by the demagnetizing field, explains why the microstructural parameter $\alpha$ is smaller than 1 for perfectly oriented grains without any defects.

We can include a soft magnetic phase at the surface of the grains in the embedded Stoner Wohlfarth method. We replace (5) with an analytic formula that gives the angular dependence of the switching fields of particles with a soft magnetic defect [18]. To test the method, we compare the results of the simplified model with conventional micromagnetics. Figure 5 shows the coercive field as function of the width of a soft magnetic phase surrounding a $Nd_2Fe_{14}B$ cube with an edge length of 200 nm. Again, there is good agreement between the simplified model and conventional micromagnetics.

The embedded Stoner-Wohlfarth model creates a map of the local switching field within a magnet. Encouraged by the good agreement between the embedded Stoner Wohlfarth model and conventional micromagnetics we now apply the embedded Stoner Wohlfarth model to a large grained $Nd_2Fe_{14}B$ magnet. The magnet has a size of 8×8×8 (µm)³ and consists of 512 grains. The grains are separated by a 2 nm thick soft magnetic grain boundary phase with $K_1 = 0$. For each virtual Stoner-Wohlfarth particle the switching field is computed. The switching field is mapped by color onto the grain surfaces. The resulting plot, shown in Figure 6, gives the distribution of the local switching field within the grain structure. Magnetization reversal will be initiated where $H_{sw}$ is lowest.

In the above example the weakest points of the magnet are next to the grain boundaries. This may change by grain boundary diffusion of Dy. As shown above the $(Nd,Dy)_2Fe_{14}B$ shell around the $Nd_2Fe_{14}B$ core of the grain improves coercivity. This change of coercivity is associated with a shift of the spot at which the magnetization reversal starts. The magnetization configuration at the saddle point shows the onset of thermally induced magnetization reversal. At the saddle point the magnetization rotates out of the anisotropy direction by around 90 degrees. Thus, it can be visualized. The spot where magnetization starts is shown in Figure 7 for the three different levels of Dy diffusion. All three samples in Figure 7 have the same grain structure. The only difference is the level of Dy diffusion. The grain structure and the diffusion levels for the 3 simulations are given in Figure 3. The $Nd_2Fe_{14}B$ core has the following material parameters $K_1$ = 4.3 MJ/m³, $\mu_0 M_s$ = 1.61 T, and $A$ = 7.7 pJ/m. The magnetization and exchange constant of the 3 nm thick grain boundary phase are $\mu_0 M_s$ = 0.5 T and $A$ = 7.7 pJ/m. The material parameters of the $(Nd,Dy)_2Fe_{14}B$ shell are $K_1$ = 5.17 MJ/m³, $\mu_0 M_s$ = 1.15 T and $A$ = 8.7 pJ/m.

In the magnet without Dy magnetization reversal is initiated at a grain boundary at the outer edge of the magnet. The location of the saddle point is shown in the image on the left-hand side of Figure 7. When the Dy containing shell covers the outer grains the weakest spot shifts inside. The shell has a thickness of 5 nm and covers only the grains which are located at the surface of the magnet. Magnetization reversal starts at a grain boundary junction in the middle of the magnet. The image in the center of Figure 7 shows a slice through the sample. In the 3$^{rd}$ simulation the Dy containing shell covers all grains. As a result, the grain surfaces are no more weak points. The Dy containing shell for the grains located at the magnet's surface has a thickness of 10 nm. For the other grain its thickness is 5 nm. Nucleation of reversed domain starts inside a grain. In order to visualize the nucleus, we again show a slice through the magnet given in the image on the right-hand side of Figure 7. Please note that the grain structure for all three scenarios is the same. However, in order to capture the initial nucleus, the slice for visualization is taken at different positions.

## 4. Conclusion

Micromagnetic simulation take into account the detailed shape and morphology of grains and intergranular phases as well as the local variation of the intrinsic magnetic properties depending on chemical compositions. Using detailed micromagnetic simulations and model order reduction, we can quantify the reduction of the coercive field owing to ferromagnetic boundary phases, demagnetizing fields, and thermal fluctuations. Successful hardening of magnets by grain boundary engineering shifts the spot where reversed domains are formed from the grain boundary junction to the center of the grains.

**Acknowledgement**

Work supported by the Austrian Science Fund (FWF): F4112 SFB ViCoM and Japan Science and the Technology Agency (JST): CREST.

Figure 1: Grain structures used for computing the microstructural parameters as function of the shape of the grains. The volume fraction of the grain boundary phase is 26 percent.

Figure 2: Comparison of conventional micromagnetics and the embedded Stoner Wohlfarth method. *Top:* Magnetization configuration obtained from conventional micromagnetics for perfect alignment and a grain size of 90 nm. Top: $\mu_0 H_{ext}$ = -5.93 T. Bottom: $\mu_0 H_{ext}$ = -5.96 T. *Bottom:* Coercive field as function of the grain size. The dashed blue lines are the results of the embedded Stoner-Wohlfarth method; the orange symbols are the results of conventional micromagnetics. Full symbols refer to zero misalignment; the open symbols give the result with the field rotated 8° off the anisotropy axes. The inset shows a map of the local Stoner-Wohlfarth switching field.

Figure 3: Model magnet for the analysis of thermal activation. *Left:* Granular structure. *Right:* Slice through the finite element mesh. The grains are separated by a 3 nm weakly ferromagnetic grain boundary phase. The thickness of the $(Nd,Dy)_2Fe_{14}B$ varies depending on the sample.

Figure 4: Reduction of coercive field in Dy diffused permanent magnets. Left: No Dy diffusion, Center: Dy covers only surface grains. Right: Dy covers all grains.

Figure 5: Comparison of conventional micromagnetics (squares) and a simplified analytic model (triangle) for the reversal of a $Nd_2Fe_{14}B$ cube with a soft magnetic surface layer of varying thickness.

Figure 6: Map of the local switching field for a large grained $Nd_2Fe_{14}B$ magnet for *h* = 0. The color map gives the local switching field computed by the embedded Stoner-Wohlfarth model. Magnetization reversal will start where the local switching field has its lowest value.

Figure 7: The initially reversed nucleus corresponds to the lowest saddle point between the remanent state and the reversed state. At the saddle point the magnetization rotates out of the anisotropy axes.

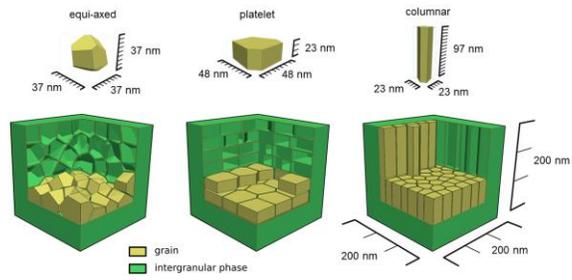

Figure 1

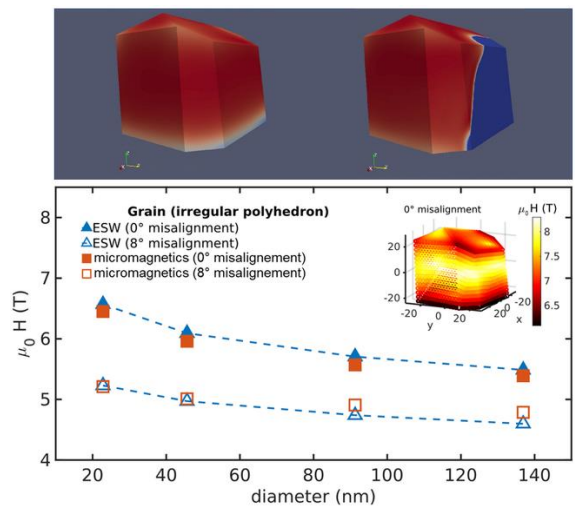

Figure 2

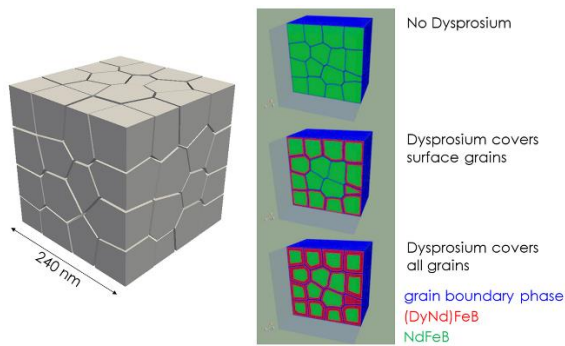

Figure 3

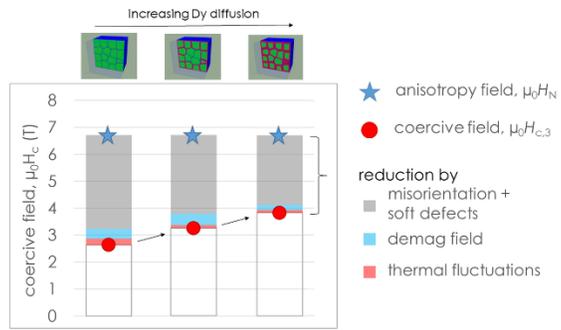

Figure 4

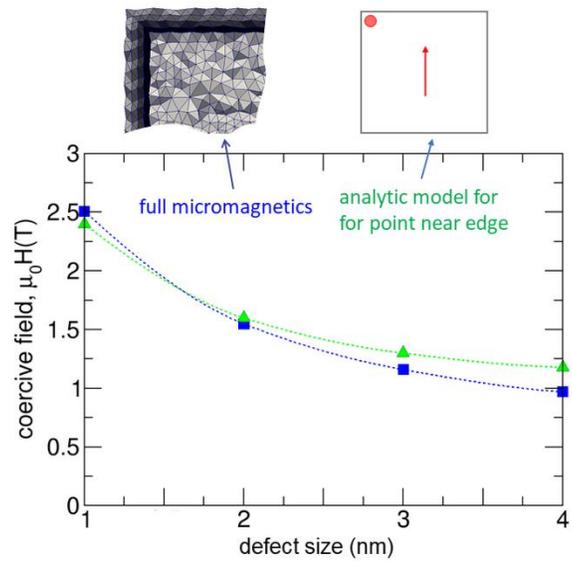

Figure 5

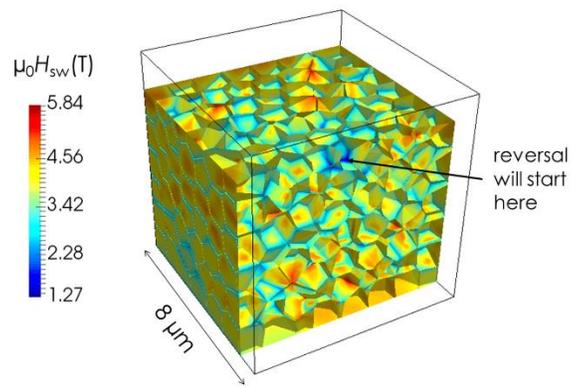

Figure 6

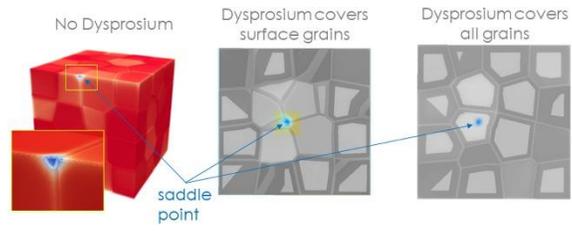

Figure 7